\documentclass[pre,twocolumn,aps,showpacs]{revtex4}
\newcommand{\bec}[1]{\mbox{\boldmath $ #1$}}
\usepackage{graphicx}
\begin{document}
\title{Generation of large-scale vorticity in a homogeneous
turbulence\\ with a mean velocity shear}
\author{Tov Elperin}
\email{elperin@menix.bgu.ac.il}
\homepage{http://www.bgu.ac.il/~elperin}
\author{Nathan Kleeorin}
\email{nat@menix.bgu.ac.il}
\author{Igor Rogachevskii}
\email{gary@menix.bgu.ac.il}
\homepage{http://www.bgu.ac.il/~gary}
\affiliation{The Pearlstone Center for Aeronautical Engineering
Studies, Department of Mechanical Engineering, Ben-Gurion
University of the Negev, Beer-Sheva 84105, P. O. Box 653, Israel}
\date{\today}
\begin{abstract}
An effect of a mean velocity shear on a turbulence and on the
effective force which is determined by the gradient of Reynolds
stresses is studied. Generation of a mean vorticity in a
homogeneous incompressible turbulent flow with an imposed mean
velocity shear due to an excitation of a large-scale instability
is found. The instability is caused by a combined effect of the
large-scale shear motions (''skew-induced" deflection of
equilibrium mean vorticity) and ''Reynolds stress-induced"
generation of perturbations of mean vorticity. Spatial
characteristics, such as the minimum size of the growing
perturbations and the size of perturbations with the maximum
growth rate, are determined. This instability and the dynamics of
the mean vorticity are associated with the Prandtl's turbulent
secondary flows. This instability is similar to the mean-field
magnetic dynamo instability. Astrophysical applications of the
obtained results are discussed.
\end{abstract}


\maketitle

\section{Introduction}

Vorticity generation in turbulent and laminar flows was studied
experimentally, theoretically and numerically in a number of
publications (see, e.g.,
\cite{P52,T56,B87,BB64,P70,M84,GHW02,PA02,C94,T98,RAO98,P87,GLM97}).
For instance, a mechanism of the vorticity production in laminar
compressible fluid flows consists in the misalignment of pressure
and density gradients \cite{P87,GLM97}. It was shown in
\cite{GLM97} that the vorticity generation represents a generic
property of any slow nonadiabatic laminar gas flow. In
incompressible flows this effect does not occur. The role of
small-scale vorticity production in incompressible turbulent flows
was discussed in \cite{T98}.

On the other hand, generation and dynamics of the mean vorticity
are associated with turbulent secondary flows (see, e.g.,
\cite{P52,B87,BB64,P70,M84,GHW02,PA02}). These flows, e.g., arise
at the lateral boundaries of three-dimensional thin shear layers
whereby longitudinal (streamwise) mean vorticity is generated by a
lateral deflection or ''skewing" of an existing shear layer
\cite{B87}. The skew-induced streamwise mean vorticity generation
corresponds to Prandtl's first kind of secondary flows. In
turbulent flows, e.g., in straight noncircular ducts, streamwise
mean vorticity can be generated by the Reynolds stresses. The
latter is Prandtl's second kind of turbulent secondary flows, and
it ''has no counterpart in laminar flow and cannot be described by
any turbulence model with an isotropic eddy viscocity" \cite{B87}.

   In the present study we demonstrated that in a homogeneous
incompressible turbulent flow with an imposed mean velocity shear
a large-scale instability can be excited which results in a mean
vorticity production. This instability is caused by a {\em
combined} effect of the large-scale shear motions (''skew-induced"
deflection of equilibrium mean vorticity) and ''Reynolds
stress-induced" generation of perturbations of mean vorticity. The
''skew-induced" deflection of equilibrium mean vorticity $
\bar{\bf W}^{(s)} $ is determined by $ (\bar{\bf W}^{(s)} \cdot
\bec{\nabla}) \tilde{\bf U} $-term in the equation for the mean
vorticity, where $ \tilde{\bf U} $ are perturbations of the mean
velocity (see below). The ''Reynolds stress-induced" generation of
a mean vorticity is determined by $ \bec{\nabla} {\bf \times}
\bec{\cal F} ,$ where $ \bec{\cal F} $ is an effective force
caused by a gradient of Reynolds stresses.

This instability is similar to the mean-field magnetic dynamo
instability (see, e.g., \cite{M78}) which is caused by a combined
effect of a nonuniform mean flow (differential rotation or
large-scale shear motions) and turbulence effects (helical
turbulent motions which produce the $\alpha$ effect \cite{M78} or
anisotropic turbulent motions which cause the ''shear-current"
effect \cite{RK03}).

This paper is organized as follows. In Section II the governing
equations are formulated. In Section III the general form of the
Reynolds stresses in a homogeneous turbulence with an imposed mean
velocity shear is found using simple symmetry reasoning, and the
mechanism for the large-scale instability caused by a combined
effect of the large-scale shear motions and ''Reynolds
stress-induced" generation of perturbations of the mean vorticity
is discussed. In Section IV the equation for the second moment of
velocity fluctuations in a homogeneous turbulence with an imposed
mean velocity shear is derived. This allows us to study an effect
of a mean velocity shear on a turbulence and to calculate the
effective force determined by the gradient of Reynolds stresses.
Using the derived mean-field equation for vorticity we studied in
Section IV the large-scale instability which causes the mean
vorticity production.

\section{The governing equations}

Our goal is to study an effect of mean velocity shear on a
turbulence and on a dynamics of a mean vorticity. The equation for
the evolution of vorticity $ {\bf W} \equiv \bec{\nabla} {\bf
\times} {\bf v} $ reads
\begin{eqnarray}
{\partial {\bf W} \over \partial t} = \bec{\nabla} {\bf \times}
({\bf v} {\bf \times} {\bf W} - \nu \bec{\nabla} {\bf \times} {\bf
W}) \;, \label{B1}
\end{eqnarray}
where ${\bf v}$ is the fluid velocity with $\bec{\nabla} \cdot
{\bf v} = 0 $ and $ \nu$ is the kinematic viscosity. This equation
follows from the Navier-Stokes equation. In this study we use a
mean field approach whereby the velocity and vorticity are
separated into the mean and fluctuating parts: $ {\bf v} =
\bar{\bf U} + {\bf u} $ and $ {\bf W} = \bar{\bf W} + {\bf w} ,$
the fluctuating parts have zero mean values, and $ \bar{\bf U} =
\langle {\bf v} \rangle ,$ $ \, \bar{{\bf W}} = \langle {\bf W}
\rangle .$ Averaging Eq.~(\ref{B1}) over an ensemble of
fluctuations we obtain an equation for the mean vorticity
$\bar{{\bf W}}$
\begin{eqnarray}
{\partial \bar{{\bf W}} \over \partial t} = \bec{\nabla} {\bf
\times} (\bar{\bf U} {\bf \times} \bar{{\bf W}} + \langle {\bf u}
{\bf \times} {\bf w} \rangle - \nu \bec{\nabla} {\bf \times}
\bar{{\bf W}})  \; .
\label{B2}
\end{eqnarray}
Note that the effect of turbulence on the mean vorticity is
determined by the Reynolds stresses $\langle u_i u_j \rangle ,$
because
\begin{eqnarray}
\langle {\bf u} {\bf \times} {\bf w} \rangle_i = - \nabla_j
\langle u_i u_j \rangle + {1 \over 2} \nabla_i \langle {\bf u}^2
\rangle \;,
\label{B3}
\end{eqnarray}
and {\bf curl} of the last term in Eq.~(\ref{B3}) vanishes.

We consider a turbulent flow with an imposed mean velocity shear
$\nabla_i \bar{\bf U}^{(s)} ,$ where $\bar{\bf U}^{(s)}$ is a
steady state solution of the Navier-Stokes equation for the mean
velocity field.  In order to study a stability of this equilibrium
we consider perturbations $\tilde{\bf U}$ of the mean velocity,
i.e., the total mean velocity is $\bar{\bf U} = \bar{\bf U}^{(s)}
+ \tilde{\bf U} .$ Similarly, the total mean vorticity is
$\bar{\bf W} = \bar{\bf W}^{(s)} + \tilde{\bf W} ,$ where
$\bar{\bf W}^{(s)} = \bec{\nabla} {\bf \times} \bar{\bf U}^{(s)} $
and $\tilde{\bf W} = \bec{\nabla} {\bf \times} \tilde{\bf U} .$
Thus, the linearized equation for the small perturbations of the
mean vorticity, $\tilde{{\bf W}} = \bar{{\bf W}} - \bar{\bf
W}^{(s)} ,$ is given by
\begin{eqnarray}
{\partial \tilde{{\bf W}} \over \partial t} = \bec{\nabla} {\bf
\times} (\bar{\bf U}^{(s)} {\bf \times} \tilde{{\bf W}} +
\tilde{\bf U} {\bf \times} \bar{\bf W}^{(s)} + \bec{\cal F}  - \nu
\bec{\nabla} {\bf \times} \tilde{{\bf W}})  \;, \label{B4}
\end{eqnarray}
where $ {\cal F}_i = - \nabla_j [f_{ij}(\bar{\bf U}) -
f_{ij}(\bar{\bf U}^{(s)})] $ is the effective force and $ f_{ij} =
\langle u_i u_j \rangle .$ Equation~(\ref{B4}) is derived by
subtracting Eq.~(\ref{B2}) written for $\bar{\bf W}^{(s)}$ from
the corresponding equation for the mean vorticity $\bar{{\bf W}}
.$ In order to obtain a closed system of equations in Section IV
we derived an equation for the effective force $ \bec{\cal F} .$
Equation~(\ref{B4}) determines the dynamics of perturbations of
the mean vorticity. In the next Sections we will show that under
certain conditions the large-scale instability can be excited
which causes the mean vorticity production.

\section{The qualitative description}

In this Section we discuss the mechanism of the large-scale
instability. The mean velocity shear can affect a turbulence. The
reason is that additional strongly anisotropic velocity
fluctuations can be generated by tangling of the mean-velocity
gradients with the Kolmogorov-type turbulence (see FIG. 1). The
source of energy of this "tangling turbulence" is the energy of
the Kolmogorov turbulence \cite{EKRZ02}. The tangling turbulence
is an universal phenomenon, e.g., it was introduced by Wheelon
\cite{W57} and Batchelor et al. \cite{BH59} for a passive scalar
and by Golitsyn \cite{G60} and Moffatt \cite{M61} for a passive
vector (magnetic field). Anisotropic fluctuations of a passive
scalar (e.g., the number density of particles or temperature) are
produced by tangling of gradients of the mean passive scalar field
with a random velocity field. Similarly, anisotropic magnetic
fluctuations are generated by tangling of the mean magnetic field
with the velocity fluctuations. The Reynolds stresses in a
turbulent flow with a mean velocity shear is another example of a
tangling turbulence. Indeed, they are strongly anisotropic in the
presence of shear and have a steeper spectrum $ (\propto k^{-7/3})
$ than a Kolmogorov turbulence (see, e.g.,
\cite{L67,WC72,SV94,IY02,EKRZ02}). The anisotropic velocity
fluctuations of tangling turbulence were studied first by Lumley
\cite{L67}.

\begin{figure}
\centering
\includegraphics[width=8cm]{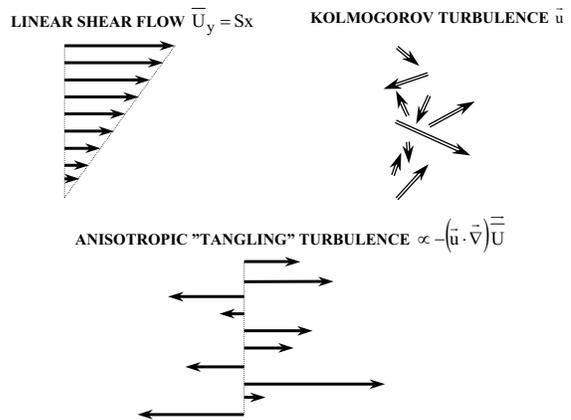}
\caption{\label{FIG1} Mechanism for the tangling turbulence:
generation of anisotropic velocity fluctuations by tangling of the
mean velocity gradients with the Kolmogorov-type turbulence.}
\end{figure}

The general form of the Reynolds stresses in a turbulent flow with
a mean velocity shear can be obtained from simple symmetry
reasoning. Indeed, the Reynolds stresses $\langle u_i u_j \rangle$
is a symmetric true tensor. In a turbulent flow with an imposed
mean velocity shear, the Reynolds stresses depend on the true
tensor $\nabla_j \bar U_{i} ,$ which can be written as a sum of
the symmetric and antisymmetric parts, i.e., $ \nabla_j \bar U_{i}
= (\partial \hat U)_{ij} - (1/2) \varepsilon_{ijk} \, \bar W_{k}
,$ where $ (\partial \hat U)_{ij} = (\nabla_i \bar U_{j} +
\nabla_j \bar U_{i}) / 2 $ is the true tensor and $\bar{\bf W} =
\bec{\nabla} {\bf \times} \bar {\bf U}$ is the mean vorticity
(pseudo-vector). We take into account the effect which is linear
in perturbations $(\partial \tilde U)_{ij}$ and $\tilde{\bf W} ,$
where $ (\partial \tilde U)_{ij} = (\nabla_i \tilde U_{j} +
\nabla_j \tilde U_{i}) / 2 .$ Thus, the general form of the
Reynolds stresses can be found using the following true tensors:
$(\partial \tilde U)_{ij} ,$ $\, M_{ij},$ $\, N_{ij} ,$ $\,
H_{ij}$ and $G_{ij} ,$ where
\begin{eqnarray}
M_{ij} &=& (\partial \bar{U}^{(s)})_{im} ({\partial \tilde
U})_{mj} + (\partial \bar{U}^{(s)})_{jm} ({\partial \tilde
U})_{mi} \;,
\label{B16}\\
N_{ij} &=& \tilde{W}_n [\varepsilon_{nim} (\partial
\bar{U}^{(s)})_{mj} + \varepsilon_{njm} (\partial
\bar{U}^{(s)})_{mi}] \;,
\label{BB17}\\
H_{ij} &=& \bar{W}^{(s)}_n [\varepsilon_{nim} (\partial \tilde
U)_{mj} + \varepsilon_{njm} (\partial \tilde U)_{mi}]  \;,
\label{B17}\\
G_{ij} &=& \bar{W}^{(s)}_i \tilde{W}_j + \bar{W}^{(s)}_j
\tilde{W}_i \;,
\label{B18}
\end{eqnarray}
$ (\partial \bar{U}^{(s)})_{ij} = (\nabla_i \bar{U}^{(s)}_{j} +
\nabla_j \bar{U}^{(s)}_{i}) / 2 $ and $\varepsilon_{ijk}$ is the
fully antisymmetric Levi-Civita tensor (pseudo-tensor). Therefore,
the Reynolds stresses have the following general form:
\begin{eqnarray}
f_{ij}(\tilde{\bf U}) &=& - 2 \nu_{_{T}} \, (\partial \tilde
U)_{ij} - l_0^2 \, (B_1 \, M_{ij}
\nonumber\\
&& + B_2 \, N_{ij} + B_3 \, H_{ij} + B_4 \, G_{ij}) \;,
\label{QB15}
\end{eqnarray}
where $ B_k $ are the unknown coefficients, $l_0$ is the maximum
scale of turbulent motions, $\nu_{_{T}} = l_0 u_0 /\beta$ is the
turbulent viscosity with the factor $ \, \beta \approx 3 - 6 ,$
and $u_0$ is the characteristic turbulent velocity in the maximum
scale of turbulent motions $l_0 .$ The parameter $l_0^2$ in
Eq.~(\ref{QB15}) was introduced using dimensional arguments. The
first term in RHS of Eq.~(\ref{QB15}) describes the standard
isotropic turbulent viscosity, whereas other terms are determined
by fluctuations caused by the imposed velocity shear $\nabla_i
\bar{U}^{(s)}_{j} .$

Let us study the evolution of the mean vorticity using
Eqs.~(\ref{B4}) and~(\ref{QB15}), where ${\cal F}_i = - \nabla_j
f_{ij}(\tilde{\bf U}) $ is the effective force. We consider a
homogeneous divergence-free turbulence with a mean velocity shear,
e.g., $ \bar{\bf U}^{(s)} = (0, Sx, 0) $  and $ \bar{\bf W}^{(s)}
= (0,0,S) .$ For simplicity we use perturbations of the mean
vorticity in the form $ \tilde{{\bf W}} = (\tilde{W}_x(z),
\tilde{W}_y(z), 0) .$ Then Eq.~(\ref{B4}) can be written as
\begin{eqnarray}
{\partial \tilde{W}_x \over \partial t} &=& S \tilde{W}_y +
\nu_{_{T}} \tilde{W}''_x  \;,
\label{E2}\\
{\partial \tilde{W}_y \over \partial t} &=& - \alpha \, S \, l_0^2
\, \tilde{W}''_x + \nu_{_{T}} \tilde{W}''_y  \;, \label{E3}
\end{eqnarray}
where $ \tilde{W}''_x = \partial^2 \tilde{W}_x /
\partial z^2 ,$ $ \, \alpha = (B_1 + 2(B_2 + B_3) - 4 B_4) / 4 .$
In Eq.~(\ref{E2}) we took into account that $ l_0^2 \tilde{W}''_y
\ll \tilde{W}_y ,$ i.e., the characteristic scale $ L_W $ of the
mean vorticity variations is much larger than the maximum scale of
turbulent motions $ l_0 .$ This assumption corresponds to the
mean-field approach. For derivation of Eqs.~(\ref{E2})
and~(\ref{E3}) we used the identities presented in Appendix A.

We seek for a solution of Eqs.~(\ref{E2}) and~(\ref{E3}) in the
form $ \propto \exp(\gamma t + i K z) .$ Thus, when $\alpha > 0$
perturbations of the mean vorticity can grow in time and the
growth rate of the instability is given by
\begin{eqnarray}
\gamma = \sqrt{\alpha} \, S \, l_0 \, K - \nu_{_{T}} \, K^2 \; .
\label{E4}
\end{eqnarray}
The maximum growth rate of perturbations of the mean vorticity, $
\gamma_{\rm max} = \alpha \, (S \, l_0)^2 / 4 \nu_{_{T}} ,$ is
attained at $ K = K_m = \sqrt{\alpha} \, S \, l_0 /2 \nu_{_{T}} .$
The sufficient condition $ \gamma > 0 $ for the excitation of the
instability reads $ L_W / l_0 > 2 \pi / (\beta \, \sqrt{\alpha} \,
\tau_0 S) ,$ where $ L_W \equiv 2 \pi / K $ and we consider a weak
velocity shear $\tau_0 S \ll 1 .$

Now let us discuss the mechanism of this instability using a
terminology from \cite{B87}. The first term, $S \tilde{W}_y =
(\bar{\bf W}^{(s)}~\cdot~\bec{\nabla})~\tilde{U}_x ,$ in
Eq.~(\ref{E2}) describes a ''skew-induced" generation of
perturbations of the mean vorticity $ \tilde{W}_x $ by
quasi-inviscid deflection of the equilibrium mean vorticity
$\bar{\bf W}^{(s)} .$ In particular, the mean vorticity $
\tilde{W}_x $ is generated from $ \tilde{W}_y $ by equilibrium
shear motions with the mean vorticity $\bar{\bf W}^{(s)} ,$ i.e.,
$ \tilde{W}_x {\bf e}_x \propto (\bar{\bf W}^{(s)} \cdot
\bec{\nabla}) \tilde{U}_x {\bf e}_x \propto \tilde{W}_y {\bf e}_x
\times \bar{\bf W}^{(s)} $ (see FIG. 2). Here ${\bf e}_x$ and
${\bf e}_y$ are the unit vectors along $x$ and $y$ axis. On the
other hand, the first term, $- \alpha \, S \, l_0^2 \,
\tilde{W}''_x ,$ in Eq.~(\ref{E3}) determines a ''Reynolds
stress-induced" generation of perturbations of the mean vorticity
$ \tilde{W}_y $ by turbulent Reynolds stresses (see FIG. 3).  In
particular, this term is determined by $ (\bec{\nabla} {\bf
\times} \bec{\cal F})_y ,$ where $ \bec{\cal F} $ is a gradient of
Reynolds stresses. This implies that the mean vorticity $
\tilde{W}_y $ is generated by an effective anisotropic viscous
term $ \propto -  l_0^2 \, S \, \tilde{W}''_x $ which is due to
the equilibrium shear motions. The growth rate of this instability
is caused by a combined effect of the sheared motions
(''skew-induced" generation) and the ''Reynolds stress-induced"
generation of perturbations of the mean vorticity.

\begin{figure}
\centering
\includegraphics[width=8cm]{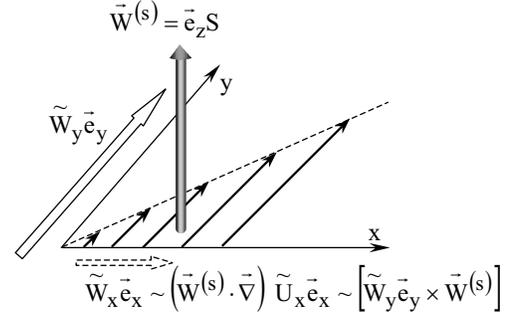}
\caption{\label{FIG2} Mechanism for ''skew-induced" generation of
perturbations of the mean vorticity $ \tilde{W}_x $ by
quasi-inviscid deflection of the equilibrium mean vorticity
$\bar{\bf W}^{(s)} $. In particular, the mean vorticity $
\tilde{W}_x $ is generated by an interaction of perturbations of
the mean vorticity $ \tilde{W}_y $ and the equilibrium mean
vorticity $\bar{\bf W}^{(s)} ,$ i.e., $ \tilde{W}_x {\bf e}_x
\propto (\bar{\bf W}^{(s)} \cdot \bec{\nabla}) \tilde{U}_x {\bf
e}_x \propto \tilde{W}_y {\bf e}_y \times \bar{\bf W}^{(s)} .$}
\end{figure}

\begin{figure}
\centering
\includegraphics[width=8cm]{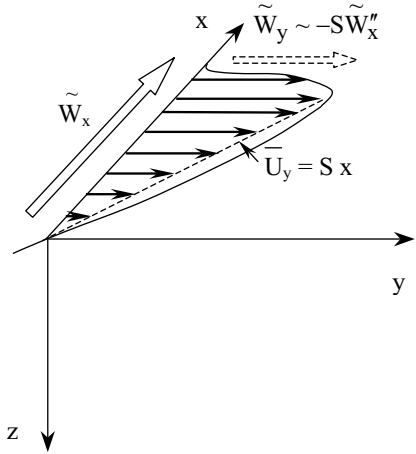}
\caption{\label{FIG3} Mechanism for a ''Reynolds stress-induced"
generation of perturbations of the mean vorticity $ \tilde{W}_y $
by turbulent Reynolds stresses. In particular, the mean vorticity
$ \tilde{W}_y $ is generated by an effective anisotropic viscous
term $ \propto -  l_0^2 \, S \, \tilde{W}''_x $ which is caused by
the equilibrium shear motions.}
\end{figure}

This large-scale instability is similar to a mean-field magnetic
dynamo instability (see, e.g., \cite{M78}). Indeed, the first term
in Eq.~(\ref{E2}) is similar to the differential rotation (or
large-scale shear motions) which causes a generation of a toroidal
mean magnetic field by a streching of the poloidal mean magnetic
field with the differential rotation. On the other hand, the first
term in Eq.~(\ref{E3}) is similar to the $\alpha$ effect
\cite{M78}, or to the ''shear-current" effect \cite{RK03}. These
effects result in generation of a poloidal mean magnetic field
from the toroidal mean magnetic field. The $\alpha$-effect is
related with the hydrodynamic helicity of an inhomogeneous
turbulent flow, while the ''shear-current" effect occurs due to an
interaction of the mean vorticity and electric current in a
homogeneous anisotropic turbulent flow of a conducting fluid. The
magnetic dynamo instability is a combined effect of nonuniform
mean flow (differential rotation or large-scale shear motions) and
turbulence effects (helical turbulent motions which produce the
$\alpha$ effect or anisotropic turbulent motions which cause
''shear-current" effect).

On the other hand, the magnetic dynamo instability is different
from the instability of the mean vorticity although they are
governed by similar equations. The mean vorticity $ \bar{\bf W} =
\bec{\nabla} {\bf \times} \bar{\bf U} $ is directly determined by
the velocity field $ \tilde{\bf U} ,$ while the magnetic field
depends on the velocity field through the induction and
Navier-Stokes equations.

\section{Effect of a mean velocity shear on a turbulence
and large-scale instability}

In this section we study quantitatively an effect of a mean
velocity shear on a turbulence. This allows us to derive an
equation for the effective force $ \bec{\cal F} $ and to study the
dynamics of the mean vorticity.

\subsection{Method of derivations}

To study an effect of a mean velocity shear on a turbulence we
used equation for fluctuations $ {\bf u}(t,{\bf r}) $ which is
obtained by subtracting equation for the mean field from the
corresponding equation for the total field:
\begin{eqnarray}
{\partial {\bf u} \over \partial t} &=& - (\bar{\bf U} \cdot
\bec{\nabla}) {\bf u} - ({\bf u} \cdot \bec{\nabla}) \bar{\bf U} -
{\bec{\nabla} p \over \rho}
\nonumber\\
& & + {\bf F}^{\rm (st)} + {\bf U}_{N} \;,
\label{B5}
\end{eqnarray}
where $p$ are the pressure fluctuations, $\rho$ is the fluid
density, ${\bf F}^{\rm (st)}$ is an external stirring force with a
zero mean value, and $ U_{N} = \langle ({\bf u} \cdot
\bec{\nabla}) {\bf u} \rangle - ({\bf u} \cdot \bec{\nabla}) {\bf
u} + \nu \Delta {\bf u} .$ We consider a turbulent flow with large
Reynolds numbers $ ({\rm Re} = l_{0} u_{0} / \nu \gg 1). $ We
assumed that there is a separation of scales, i.e., the maximum
scale of turbulent motions $ l_0 $ is much smaller then the
characteristic scale of inhomogeneities of the mean fields. Using
Eq.~(\ref{B5}) we derived equation for the second moment of
turbulent velocity field $ f_{ij}({\bf k, R}) \equiv \int \langle
u_i ({\bf k} + {\bf K} / 2) u_j(-{\bf k} + {\bf K} / 2) \rangle
\exp(i {\bf K} {\bf \cdot} {\bf R}) \,d {\bf K} $:
\begin{eqnarray}
{\partial f_{ij}({\bf k, R}) \over \partial t} = I_{ijmn}(\bar{\bf
U}) f_{mn} + F_{ij} + f_{ij}^{(N)} \;
\label{A6}
\end{eqnarray}
(see Appendix B), where
\begin{eqnarray}
I_{ijmn}(\bar{\bf U}) &=& \biggl(2 k_{iq} \delta_{mp} \delta_{jn}
+ 2 k_{jq} \delta_{im} \delta_{pn} - \delta_{im} \delta_{jq}
\delta_{np}
\nonumber\\
&& - \delta_{iq} \delta_{jn} \delta_{mp} + \delta_{im} \delta_{jn}
k_{q} {\partial \over \partial k_{p}} \biggr) \nabla_{p} \bar
U_{q} \;,
\label{A14}
\end{eqnarray}
and $ {\bf R} $ and $ {\bf K} $ correspond to the large scales,
and $ {\bf r} $ and $ {\bf k} $ to the small scales (see Appendix
B), $ k_{ij} = k_i  k_j / k^2 ,$ $\, f_{ij}^{(N)}({\bf k},{\bf R})
$ is the third moment appearing due to the nonlinear term, $
\bec{\nabla} =
\partial / \partial {\bf R} ,$ $ F_{ij}({\bf k},{\bf R}) = \langle
\tilde F_i ({\bf k},{\bf R}) u_j(-{\bf k},{\bf R}) \rangle +
\langle u_i({\bf k},{\bf R}) \tilde F_j(-{\bf k },{\bf R}) \rangle
$ and $ {\bf \tilde F} ({\bf k},{\bf R},t) = - {\bf k} {\bf
\times} ({\bf k} {\bf \times} {\bf F}^{\rm (st)} ({\bf k},{\bf
R})) / k^2 .$ Equation~(\ref{A6}) is written in a frame moving
with a local velocity $ \bar {\bf U} $ of the mean flow. In
Eqs.~(\ref{A6}) and (\ref{A14}) we neglected small terms which are
of the order of $O(|\nabla^3 \bar{\bf U}|) .$ Note that
Eqs.~(\ref{A6}) and (\ref{A14}) do not contain terms proportional
to $O(|\nabla^2 \bar{\bf U}|) .$

The total mean velocity is $\bar{\bf U} = \bar{\bf U}^{(s)} +
\tilde{\bf U} ,$ where we considered a turbulent flow with an
imposed mean velocity shear $\nabla_i \bar{\bf U}^{(s)} .$ Now let
us introduce a background turbulence with zero gradients of the
mean fluid velocity $ \nabla_{i} \bar U_{j} = 0 .$ The background
turbulence is determined by equation $ \partial f_{ij}^{(0)} /
\partial t = F_{ij} + f_{ij}^{(N0)} ,$ where the superscript
$ {(0)} $ corresponds to the background turbulence, and we assumed
that the tensor $F_{ij}({\bf k, R}) ,$ which is determined by a
stirring force, is independent of the mean velocity. Equation for
the deviations $f_{ij} - f_{ij}^{(0)}$ from the background
turbulence is given by
\begin{eqnarray}
{\partial (\hat f - \hat f^{(0)}) \over \partial t} = [\hat
I(\bar{\bf U}^{(s)})  + \hat I(\tilde{\bf U})] \hat f + \hat
f^{(N)} - \hat f^{(N0)} \;, \label{B8}
\end{eqnarray}
where we used the following notations: $\hat f \equiv f_{ij}({\bf
k, R}) ,$ $\, \hat f^{(N)} \equiv f_{ij}^{(N)}({\bf k, R}) ,$ $\,
\hat f^{(N0)} \equiv f_{ij}^{(N0)}({\bf k, R}) ,$ $\, \hat f^{(0)}
\equiv f_{ij}^{(0)}({\bf k, R}) ,$ and $ \hat I(\bar{\bf U}) \hat
f \equiv I_{ijmn}(\bar{\bf U}) f_{mn}({\bf k, R}) .$

Equation~(\ref{B8}) for the deviations of the second moments in
${\bf k}$-space contains the deviations of the third moments and a
problem of closing the equations for the higher moments arises.
Various approximate methods have been proposed for the solution of
problems of this type (see, e.g., \cite{MY75,O70,Mc90}). The
simplest procedure is the $ \tau $ approximation which was widely
used for study of different problems of turbulent transport (see,
e.g., \cite{O70,PFL76,KRR90,RK2000}). One of the simplest
procedures, that allows us to express the deviations of the third
moments $\hat f^{(N)} - \hat f^{(N0)}$ in ${\bf k}$-space in terms
of that for the second moments $\hat f - \hat f^{(0)} ,$ reads
\begin{eqnarray}
\hat f^{(N)} - \hat f^{(N0)} &=& - {\hat f - \hat f^{(0)} \over
\tau (k)} \;,
\label{A1}
\end{eqnarray}
where $ \tau (k) $ is the correlation time of the turbulent
velocity field. Here we assumed that the time $ \tau(k) $ is
independent of the gradients of the mean fluid velocity because in
the framework of the mean-field approach we may only consider a
weak shear: $ \tau_0 |\bec{\nabla} \bar U| \ll 1 ,$ where $ \tau_0
= l_{0} / u_{0} .$

The $ \tau $-approximation  is in general similar to Eddy Damped
Quasi Normal Markowian (EDQNM) approximation. However there is a
principle difference between these two approaches (see
\cite{O70,Mc90}). The EDQNM closures do not relax to the
equilibrium, and do not describe properly the motions in the
equilibrium state. Within the EDQNM theory, there is no
dynamically determined relaxation time, and no slightly perturbed
steady state can be approached \cite{O70}. In the $ \tau
$-approximation, the relaxation time for small departures from
equilibrium is determined by the random motions in the equilibrium
state, but not by the departure from equilibrium \cite{O70}.
Analysis performed in \cite{O70} showed that the $ \tau
$-approximation describes the relaxation to the equilibrium state
(the background turbulence) more accurately than the EDQNM
approach.

Note that we applied the $ \tau $-approximation (\ref{A1}) only to
study the deviations from the background turbulence which are
caused by the spatial derivatives of the mean velocity. The
background turbulence is assumed to be known. Here we used the
following model of the background isotropic and homogeneous
turbulence:
\begin{eqnarray}
f^{(0)}_{ij}({\bf k},{\bf R}) &=& \frac{u_0^2}{8 \pi k^{2}}
P_{ij}({\bf k}) {\cal E}(k)  \;, \label{A2}
\end{eqnarray}
where $ P_{ij}({\bf k}) = \delta_{ij} - k_{ij} ,$ $\, \delta_{ij}
$ is the Kronecker tensor, $ \tau(k) = 2 \tau_0 \bar \tau(k) ,$ $
\, {\cal E}(k) = - d \bar \tau(k) / dk ,$ $ \, \bar \tau(k) = (k /
k_{0})^{1-q} ,$ $ \, 1 < q < 3 $  is the exponent of the kinetic
energy spectrum (e.g., $ q = 5/3 $ for Kolmogorov spectrum), and $
k_{0} = 1 / l_{0} .$

\subsection{Equation for the second moment of velocity
fluctuations}

We assume that the characteristic time of variation of the second
moment $f_{ij}({\bf k},{\bf R})$ is substantially larger than the
correlation time $\tau(k)$ for all turbulence scales. Thus in a
steady-state Eq.~(\ref{B8}) reads
\begin{eqnarray}
\hat L (\hat f - \hat f^{(0)}) = \tau(k) \, [\hat I(\bar{\bf
U}^{(s)}) + \hat I(\tilde{\bf U})] \, \hat f^{(0)} \;, \label{B9}
\end{eqnarray}
where $ \hat L \equiv L_{ijmn} = \delta_{im} \delta_{jn} - \tau(k)
\, [ I_{ijmn}(\bar{\bf U}^{(s)}) + I_{ijmn}(\tilde{\bf U})] ,$ and
we used Eq.~(\ref{A1}). The solution of Eq.~(\ref{B9}) yields the
second moment $ \hat f \equiv f_{ij}({\bf k},{\bf R})$:
\begin{eqnarray}
\hat f &\approx& \hat f(\bar{\bf U}^{(s)}) + \tau(k) \, [\hat
I(\tilde{\bf U}) + \hat I(\bar{\bf U}^{(s)}) \, \tau(k) \, \hat
I(\tilde{\bf U})
\nonumber\\
&& + \hat I(\tilde{\bf U}) \, \tau(k) \, \hat I(\bar{\bf
U}^{(s)})] \, \hat f^{(0)} \;, \label{B11}
\end{eqnarray}
where $ \hat f(\bar{\bf U}^{(s)}) = \hat f^{(0)} + \tau(k) [\hat
I(\bar{\bf U}^{(s)}) + \hat I(\bar{\bf U}^{(s)}) \tau(k) \hat
I(\bar{\bf U}^{(s)})] \hat f^{(0)} .$ In Eq.~(\ref{B11}) we
neglected terms which are of the order of $O(|\nabla \tilde{\bf
U}|^2) $ and $O(\, |\nabla \bar{\bf U}^{(s)}|^2 |\nabla \tilde{\bf
U}|) .$ The first term in the equation for $\hat f(\bar{\bf
U}^{(s)})$ is independent of the mean velocity shear and it
describes the background turbulence, while the second and the
third terms in this equation determine an effect of the mean
velocity shear on turbulence.

\subsection{Effective force}

Equation~(\ref{B11}) allows us to determine the effective force:
${\cal F}_i = - {\nabla}_j \int \tilde f_{ij}({\bf k},{\bf R}) \,
d{\bf k} ,$ where $\tilde f = \hat f - \hat f(\bar{\bf U}^{(s)}),$
and we used notation $\tilde f \equiv \tilde f_{ij}({\bf k},{\bf
R}) .$ The integration in ${\bf k}$-space yields the second moment
$ \tilde f_{ij}({\bf R}) = \int \tilde f_{ij}({\bf k},{\bf R}) \,
d{\bf k} $:
\begin{eqnarray}
\tilde f_{ij}({\bf R}) &=& - 2 \nu_{_{T}} \, (\partial \tilde
U)_{ij} - l_0^2 \, [4 C_1 \, M_{ij}
\nonumber\\
&& + C_2 \, (N_{ij} + H_{ij}) + C_3 \, G_{ij}] \;,
\label{B15}
\end{eqnarray}
where $C_1 = 8 (q^2 - 13 q + 40) / 315 ,$ $ \, C_2 = 2 (6 - 7 q) /
45 ,$ $ \, C_3 = - 2 (q + 2) / 45 ,$ and the tensors $ M_{ij},$
$\, N_{ij} ,$ $\, H_{ij}$ and $G_{ij} $ are determined by Eqs.
(\ref{B16})-(\ref{B18}). In Eq.~(\ref{B15}) we omitted terms $
\propto \delta_{ij} $ because they do not contribute to $
\bec{\nabla} {\bf \times} \bec{\cal F} $ (see Eq.~(\ref{B4}) for
perturbations of the mean vorticity $\tilde {\bf W} ).$ To derive
Eq.~(\ref{B15}) we used the identities presented in Appendix A.
Equations~(\ref{QB15}) and ~(\ref{B15}) yield $B_1 = 4 C_1 ,$ $\,
B_2 = B_3 = C_2$ and $B_4 = C_3 .$

Note that the mean velocity gradient $ \nabla_i \bar{\bf U}^{(s)}
$ causes generation of anisotropic velocity fluctuations (tangling
turbulence). Inhomogeneities of perturbations of the mean velocity
$\tilde {\bf U} $ produce additional velocity fluctuations, so
that the Reynolds stresses $\tilde f_{ij}({\bf R})$ are the result
of a combined effect of two types of velocity fluctuations
produced by the tangling of mean gradients $ \nabla_i \bar{\bf
U}^{(s)} $ and $ \nabla_i \tilde{\bf U} $ by a small-scale
Kolmogorov turbulence. Equation~(\ref{B15}) allows to determine
the effective force $ {\cal F}_i = - {\nabla}_j \tilde f_{ij}({\bf
R}) .$

\subsection{The large-scale instability in a homogeneous turbulence
with a mean velocity shear}

Let us study the evolution of the mean vorticity using
Eq.~(\ref{B15}) for the Reynolds stresses. Consider a homogeneous
turbulence with a mean velocity shear, e.g., $ \bar{\bf U}^{(s)} =
(0, Sx, 0) $ and $ \bar{\bf W}^{(s)} = (0,0,S) .$ For simplicity
we consider perturbations of the mean vorticity in the form $
\tilde{{\bf W}} = (\tilde{W}_x(z), \tilde{W}_y(z), 0) .$ Then
Eq.~(\ref{B4}) reduces to Eqs.~(\ref{E2}) and~(\ref{E3}), where
$\alpha = C_1 + C_2 - C_3 = 4 (2 q^2 - 47 q + 108) / 315 ,$ and we
used Eq.~(\ref{B15}). We seek for a solution of Eqs.~(\ref{E2})
and~(\ref{E3}) in the form $ \propto \exp(\gamma t + i K z) .$
Thus, the growth rate of perturbations of the mean vorticity is
given by $ \gamma = \sqrt{\alpha} \, S \, l_0 \, K - \nu_{_{T}} \,
K^2 .$ The maximum growth rate of perturbations of the mean
vorticity, $ \gamma_{\rm max} = \alpha \, (S \, l_0)^2 / 4
\nu_{_{T}} \approx 0.1 \, \beta \, S^2 \tau_0 ,$ is attained at $
K = K_m = \sqrt{\alpha} \, S \, l_0 /2 \nu_{_{T}} .$ Here we used
that for a Kolmogorov spectrum $ (q = 5/3) $ of the background
turbulence, the factor $ \alpha \approx 0.45 .$ The sufficient
condition $ \gamma > 0 $ for the excitation of the instability
reads $ L_W / l_0 > 2 \pi / (\beta \, \sqrt{\alpha} \, \tau_0 S)
.$ Since $\tau_0 S \ll 1 $ (we considered a weak velocity shear),
the scale $L_W \gg l_0 $ and, therefore, there is a separation of
scales. The mechanism of this instability is discussed in Section
III and is associated with a combined effect of the
''skew-induced" deflection of equilibrium mean vorticity due to
the sheared motions and the ''Reynolds stress-induced" generation
of perturbations of mean vorticity.

\section{Conclusions and applications}

We discussed an effect of a mean velocity shear on a turbulence
and on the effective force which is determined by the gradient of
Reynolds stresses. We demonstrated that in a homogeneous
incompressible turbulent flow with an imposed mean velocity shear
a large-scale instability can be excited which results in a mean
vorticity production. This instability is caused by a {\em
combined} effect of the large-scale shear motions (''skew-induced"
deflection of equilibrium mean vorticity) and ''Reynolds
stress-induced" generation of perturbations of mean vorticity. We
determined the spatial characteristics, such as the minimum size
of the growing perturbations and the size of perturbations with
the maximum growth rate.

The analyzed effect of the mean vorticity production may be of
relevance in different turbulent industrial, environmental and
astrophysical flows (see, e.g.,
\cite{B87,BB64,P70,M84,GHW02,PA02,L83,BR98,EKR98,CH00}). Thus,
e.g., the suggested mechanism can be used in the analysis of the
flows associated with the Prandtl's turbulent secondary flows
(see, e.g., \cite{B87,BB64,P70,M84,GHW02,PA02}). These flows,
e.g., arise in straight noncircular ducts, at the lateral
boundaries of three-dimensional thin shear layers, etc. The simple
model considered in the present paper can mimic the flows
associated with turbulent secondary flows.

The obtained results may be also important in astrophysics, e.g.,
in extragalactic clusters and in interstellar clouds. The
extragalactic clusters are nonrotating objects with a homogeneous
turbulence in the center of a extragalactic cluster. Sheared
motions between interacting clusters can cause an excitation of
the large-scale instability which results in the mean vorticity
production and formation of large-scale vortices. Dust particles
can be trapped by these vortices to enhance agglomeration of
material and formation of particles inhomogeneities
\cite{BR98,EKR98,CH00}. The sheared motions can also occur between
interacting interstellar clouds, whereby the dynamics of the mean
vorticity is important.

\begin{acknowledgments}
This work was partially supported by The German-Israeli Project
Cooperation (DIP) administrated by the Federal Ministry of
Education and Research (BMBF) and by INTAS Program Foundation
(Grants No. 99-348 and No. 00-0309).
\end{acknowledgments}

\appendix

\section{Identities used for derivation of Eqs.~(\ref{E2}),
(\ref{E3}) and (\ref{B15})}

To derive Eqs.~(\ref{E2}) and~(\ref{E3}) we used the following
identities:
\begin{eqnarray*}
\bec{\nabla} {\bf \times} {\bf M} &=& (S K^2 / 4) (\tilde W_y,
\tilde W_x, 0)  \;,
\\
\bec{\nabla} {\bf \times} (\tilde{\bf U} {\bf \times} \bar{\bf
W}^{(s)}) &=& S (\tilde W_y, - \tilde W_x, 0) \;,
\\
\bec{\nabla} {\bf \times} (\bar{\bf U}^{(s)} {\bf \times}
\tilde{\bf W}) &=& S (0, \tilde W_x, 0)  \;,
\\
\bec{\nabla} {\bf \times} {\bf J} &=& (S K^2 / 2) (\tilde W_y,
\tilde W_x, 0)  \;,
\\
\bec{\nabla} {\bf \times} [(\bar{\bf W}^{(s)} \cdot \bec{\nabla})
\tilde{\bf W}]  &=& S K^2 (\tilde W_y, - \tilde W_x, 0)  \;,
\end{eqnarray*}
where $M_i = \nabla_{j} M_{ij} ,$ $ J_i = \nabla_{j} [ \tilde{W}_n
(\varepsilon_{nim} (\partial \bar{U}^{(s)})_{mj} +
\varepsilon_{njm} (\partial \bar{U}^{(s)})_{mi}) ] ,$ and we also
took into account that
\begin{eqnarray*}
&& \nabla_{j} G_{ij} = (\bar{\bf W}^{(s)} \cdot \bec{\nabla})
\tilde{W}_i  \;,
\\
&& \nabla_{j} [ \bar{W}^{(s)}_n (\varepsilon_{nim} (\partial
\tilde U)_{mj} + \varepsilon_{njm} (\partial \tilde U)_{mi}) ]
\\
&& = [\nabla_i (\bar{\bf W}^{(s)} \cdot \tilde{\bf W}) - (\bar{\bf
W}^{(s)} \cdot \bec{\nabla}) \tilde{W}_i] / 2 \; .
\end{eqnarray*}

To derive Eq.~(\ref{B15}) we used the following identities for the
integration over the angles in $ {\bf k} $-space:

\begin{eqnarray*}
&& \int k_{ijmn}  \,d \hat \Omega = {4 \pi \over 15} \Delta_{ijmn}
\;,
\\
&& T_{ijmnpq} \equiv \int k_{ijmnpq}  \,d \hat \Omega = {4 \pi
\over 105} (\Delta_{mnpq} \delta_{ij}
\\
&& + \Delta_{jmnq} \delta_{ip} + \Delta_{imnq} \delta_{jp} +
\Delta_{jmnp} \delta_{iq}
\\
&& + \Delta_{imnp} \delta_{jq} + \Delta_{ijmn} \delta_{pq} -
\Delta_{ijpq} \delta_{mn}) \;,
\\
&& T_{ijmnpq} (\nabla_m \bar{U}^{(s)}_n) (\nabla_p \tilde U_q) =
{4 \pi \over 105} (4 M_{ij} + \delta_{ij} M_{pp}) \;,
\end{eqnarray*}
$ k_{ijmn} = k_{i} k_{j} k_{m} k_{n} / k^{4} ,$ $ \, d \hat \Omega
= \sin \theta \,d \theta \,d \varphi ,$ $ \, k_{ijmnpq} = k_{ijmn}
k_{pq} $ and $ \Delta_{ijmn} = \delta_{ij} \delta_{mn} +
\delta_{im} \delta_{nj} + \delta_{in} \delta_{mj} .$

\section{Derivation of Eq.~(\ref{A6})}

In order to derive Eq.~(\ref{A6}) we use a two-scale approach,
i.e., a correlation function is written as follows
\begin{eqnarray*}
\langle u_i({\bf x}) u_j ({\bf  y}) \rangle &=& \int \langle u_i
({\bf k}_1) u_j ({\bf k}_2) \rangle  \exp[i({\bf  k}_1 {\bf \cdot}
{\bf x} \\
&& + {\bf k}_2 {\bf \cdot} {\bf y})] \,d{\bf k}_1 \, d{\bf k}_2
\\
&=& \int f_{ij}({\bf k, R}) \exp(i {\bf k} {\bf \cdot} {\bf r})
\,d {\bf k} \;,
\\
f_{ij}({\bf k, R}) &=& \int \langle u_i ({\bf k} + {\bf  K} / 2)
u_j(-{\bf k} + {\bf  K} / 2) \rangle
\\
&& \times \exp(i {\bf K} {\bf \cdot} {\bf R}) \,d {\bf K} \;
\end{eqnarray*}
(see, e.g., \cite{RS75,KR94}), where $ {\bf R} $ and $ {\bf K} $
correspond to the large scales, and $ {\bf r} $ and $ {\bf k} $ to
the small scales, {\em i.e.,} $ {\bf R} = ({\bf x} +  {\bf y}) / 2
,$ $ \quad {\bf r} = {\bf x} - {\bf y},$ $ \quad {\bf K} = {\bf
k}_1 + {\bf k}_2,$ $ \quad {\bf k} = ({\bf k}_1 - {\bf k}_2) / 2
.$ This implies that we assumed that there exists a separation of
scales, i.e., the maximum scale of turbulent motions $ l_0 $ is
much smaller then the characteristic scale of inhomogeneities of
the mean fields.

Now we calculate
\begin{eqnarray}
{\partial f_{ij}({\bf k}_1,{\bf k}_2) \over \partial t} &\equiv&
\langle P_{in}({\bf k}_1) {\partial u_{n}({\bf k}_1) \over
\partial t} u_j({\bf k}_2) \rangle
\nonumber\\
&& + \langle u_i({\bf k}_1) P_{jn}({\bf k}_2) {\partial u_{n}({\bf
k}_2) \over \partial t} \rangle \;, \label{A8}
\end{eqnarray}
where we multiplied equation of motion (\ref{B5}) rewritten in $
{\bf k} $-space by $ P_{ij}({\bf k}) = \delta_{ij} - k_{ij} $ in
order to exclude the pressure term from the equation of motion, $
\delta_{ij} $ is the Kronecker tensor and $ k_{ij} = k_i  k_j /
k^2 .$ Thus, the equation for $ f_{ij}({\bf k, R}) $ is given by
Eq.~(\ref{A6}).

For the derivation of Eq.~(\ref{A6}) we used the following
identity
\begin{eqnarray}
&& i k_i \int f_{ij}({\bf k} - {1 \over 2}{\bf  Q}, {\bf K} - {\bf
Q}) \bar U_{p}({\bf  Q}) \exp(i {\bf K} {\bf \cdot} {\bf R}) \,d
{\bf  K} \,d {\bf  Q}
\nonumber\\
& & = -{1 \over 2} \bar U_{p} \nabla _i f_{ij} + {1 \over 2}
f_{ij} \nabla _i \bar U_{p} -  {i \over 4} (\nabla _s \bar U_{p})
\biggl(\nabla _i {\partial f_{ij} \over \partial k_s} \biggr)
\nonumber\\
& & +  {i \over 4} \biggl( {\partial f_{ij} \over
\partial k_s} \biggr) (\nabla _s \nabla _i \bar U_{p}) \; .
\label{D1}
\end{eqnarray}
To derive Eq.~(\ref{D1}) we multiply  the equation
$\bec{\nabla}~\cdot~{\bf u} = 0 ,$ written in ${\bf k}$-space for
$u_i({\bf k}_1 - {\bf Q}) ,$ by $u_j({\bf k}_2) \bar U_{p}({\bf
Q}) \exp(i {\bf K} {\bf \cdot} {\bf R}) ,$ and integrate over
${\bf K}$ and ${\bf Q}$, and average over ensemble of velocity
fluctuations. Here ${\bf k}_1 = {\bf k} + {\bf  K} / 2$ and ${\bf
k}_2 = -{\bf k} + {\bf K} / 2 .$ This yields
\begin{eqnarray}
&& \int i \biggl(k_i + {1 \over 2} K_i - Q_i \biggr) \langle
u_i({\bf k} + {1 \over 2}{\bf K} - {\bf Q}) u_j(-{\bf k} + {1
\over 2}{\bf K}) \rangle
\nonumber\\
& & \times \bar U_{p}({\bf  Q}) \exp{(i {\bf K} {\bf \cdot} {\bf
R})} \,d {\bf  K} \,d {\bf Q} = 0  \; . \label{D2}
\end{eqnarray}
Next, we introduce new variables: $ \tilde {\bf k}_{1} = {\bf k} +
{\bf  K} / 2 - {\bf  Q} ,$ $ \tilde {\bf k}_{2} = - {\bf k} + {\bf
K} / 2 $ and $ \tilde {\bf k} = (\tilde {\bf k}_{1} - \tilde {\bf
k}_{2}) / 2 = {\bf k} - {\bf  Q} / 2,$ $ \tilde {\bf K} = \tilde
{\bf k}_{1} + \tilde {\bf k}_{2} = {\bf  K} - {\bf  Q} .$ This
allows us to rewrite Eq.~(\ref{D2}) in the form
\begin{eqnarray}
& & \int i \biggl(k_i + {1 \over 2} K_i - Q_i \biggr) f_{ij}({\bf
k} - {1 \over 2}{\bf Q}, {\bf K} - {\bf Q}) \bar U_{p}({\bf  Q})
\nonumber\\
& & \times \exp{(i {\bf K} {\bf \cdot} {\bf R})} \,d {\bf  K} \,d
{\bf Q} = 0  \; .
\label{D3}
\end{eqnarray}
Since $ |{\bf Q}| \ll |{\bf k}| $ we can use the Taylor expansion
\begin{eqnarray}
f_{ij}({\bf k} - {\bf Q}/2, {\bf  K} - {\bf  Q}) \simeq
f_{ij}({\bf k},{\bf  K} - {\bf  Q})
\nonumber\\
- \frac{1}{2} {\partial f_{ij}({\bf k},{\bf  K} - {\bf Q}) \over
\partial k_s} Q_s  + O({\bf Q}^2) \; .
\label{D4}
\end{eqnarray}
We also use the following identities:
\begin{eqnarray}
&& [f_{ij}({\bf k},{\bf R}) \bar U_{p}({\bf R})]_{\bf  K} = \int
f_{ij}({\bf k},{\bf  K} - {\bf  Q}) \bar U_{p}({\bf Q}) \,d {\bf
Q} \;,
\nonumber \\
&& \nabla_{p} [f_{ij}({\bf k},{\bf R}) \bar U_{p}({\bf R})] = \int
i K_{p} [f_{ij}({\bf k},{\bf R}) \bar U_{p}({\bf R})]_{\bf  K}
\nonumber\\
&& \times \exp{(i {\bf K} {\bf \cdot} {\bf R})} \,d {\bf  K} \; .
\label{D5}
\end{eqnarray}
Therefore, Eqs. (\ref{D3})-(\ref{D5}) yield Eq.~(\ref{D1}).

\end{document}